\documentstyle[12pt,epsfig,amssymb]{article}

\newlength{\dinwidth}
\newlength{\dinmargin}
\begin{document}

\begin{center}

{\Large \bf Parton distribution functions from the precise NNLO QCD fit}

\vspace{1cm}
\textsc{S.~I.~Alekhin}

\vspace{0.1in}
{\baselineskip=14pt Institute for High Energy Physics, 142284 Protvino, Russia}

\begin{abstract}
We report the parton distribution functions (PDFs)
determined from the NNLO QCD analysis 
of the world inclusive DIS data with account of the 
precise NNLO QCD corrections to the evolution equations kernel.
The value of strong coupling constant 
$\alpha^{\rm NNLO}_{\rm s}(M_{\rm Z})=0.1141\pm 0.0014({\rm exp.})$, 
in fair agreement with one obtained using the earlier
approximate NNLO kernel by van Neerven-Vogt.
The intermediate bosons rates calculated in the NNLO using obtained PDFs
are in agreement to the latest Run II results. 

\end{abstract}
\end{center}
\vspace*{0.5cm}

 Account of the higher-order QCD corrections for
most of high-energy processes is important 
due to the value of strong coupling constant $\alpha_{\rm s}$
is not small for a realistic kinematics.
This is also true for the deep-inelastic lepton-nucleon 
scattering (DIS) process, which provides valuable information about 
structure of nucleon. However,  
since the next-to-next-to-leading order (NNLO) corrections were completely 
calculated only lately, mostly often the analysis of the DIS data 
was performed in the next-to-leading (NLO) approximation or, as the best,
with the approximate NNLO evolution kernels derived in 
Ref.~\cite{vanNeerven:2000wp}
on the basis of calculations~\cite{Retey:2000nq,Catani:1994sq}.
With the recently calculated exact expressions for the  
NNLO evolution kernels~\cite{Moch:2004pa}
one can improve available extractions of the NNLO PDFs  
based on the approximate 
evolution kernels getting rid of the error due to kernel uncertainty.
Even so consistent extraction of the NNLO PDFs from the global fits including
the jet production data~\cite{Martin:2004ir} is still unfeasible since 
the NNLO coefficient functions for the jet production process are not 
completely calculated. In this letter we describe the NNLO PDFs
obtained from the updated analysis
of the world data on inclusive DIS process~\cite{Alekhin:2002fv},
where the NNLO coefficient functions are known and full account of the 
NNLO corrections is possible.

We use for the analysis the charged-leptons proton/deuteron data on the DIS 
cross sections collected in the SLAC-CERN-HERA 
experiments~\cite{Whitlow:1992uw} with the cuts $Q^2>2.5~{\rm GeV}^2$, 
$W>1.8~{\rm GeV}$, and $x<0.75$ 
imposed in order to reject the kinematical regions 
problematic for the perturbative QCD (pQCD) and 
where the nuclear corrections are particularly big.
The HERA data with $Q^2>300~{\rm GeV}^2$ were also cut off since they have 
marginal impact on the precision of PDFs obtained, but complicate the analysis 
due to account of the $Z$-boson contribution is required for this kinematics. 
The pQCD evolution input for the $u$-, $d$-, $s$-quarks and gluons 
at $Q^2_0=9~{\rm GeV}^2$ is 
\begin{equation}
xu_{\rm V}(x,Q_0)=\frac{2}{N^{\rm V}_{\rm u}}
x^{a_{\rm u}}(1-x)^{b_{\rm u}}(1+\gamma_2^{\rm u}x),
\end{equation}
\begin{equation}
xu_{\rm S}(x,Q_0)=\frac{A_{\rm S}}{N_{\rm S}}
\eta_{\rm u} x^{a_{\rm s}}(1-x)^{b_{\rm su}},
\end{equation}
\begin{equation}
xd_{\rm V}(x,Q_0)=\frac{1}{N^{\rm V}_{\rm d}}x^{a_{\rm d}}(1-x)^{b_{\rm d}},
\end{equation}
\begin{equation}
xd_{\rm S}(x,Q_0)=\frac{A_{\rm S}}{N^{\rm S}}x^{a_{\rm s}}(1-x)^{b_{\rm sd}},
\end{equation}
\begin{equation}
xs_{\rm S}(x,Q_0)=\frac{A_{\rm S}}{N^{\rm S}}\eta_{\rm s}
x^{a_{\rm s}}(1-x)^{(b_{\rm su}+b_{\rm sd})/2},
\end{equation}
\begin{equation}
xG(x,Q_0)=A_{\rm G}x^{a_{\rm G}}(1-x)^{b_{\rm G}}
(1+\gamma^{\rm G}_1\sqrt{x}+\gamma^{\rm G}_2 x), 
\end{equation}
where indices $V$ and $S$ correspond to the valence and sea distributions
correspondingly.
The normalization factors $N^{\rm V}_{\rm u,d}$ and $A_{\rm G}$ 
are calculated from other parameters using the fermion number and momentum 
conservation. The value of parameter $N_{\rm S}$ is defined from 
the condition that $A_{\rm S}$ gives total momentum carried by the sea quarks.
The value of $\eta_{\rm s}$ is fixed at 0.42. For the PDFs parameters 
obtained in our fit this choice provides 
the value of strange sea suppression factor equal to 0.41 at 
$Q^2_0=20~{\rm GeV}^2$, in agreement to the CCFR/NuTeV analysis of 
Ref.~\cite{Goncharov:2001qe}.
The $b$- and $c$-quarks contributions are accounted in 
the massive scheme with the $O(\alpha_{\rm s}^2)$ correction 
of Ref.~\cite{Laenen:1992xs} included.
For the lowest $Q/W$ data used in the fit 
the power corrections are important and therefore we
take into account the target-mass correction by 
Georgi-Politzer \cite{Georgi:1976ve}
and the dynamical twist-4 terms in the structure functions $F_{\rm 2,T}$
parameterized in a model-independent way 
as the piece-linear functions of $x$.

\begin{figure}
\centerline{\epsfig{file=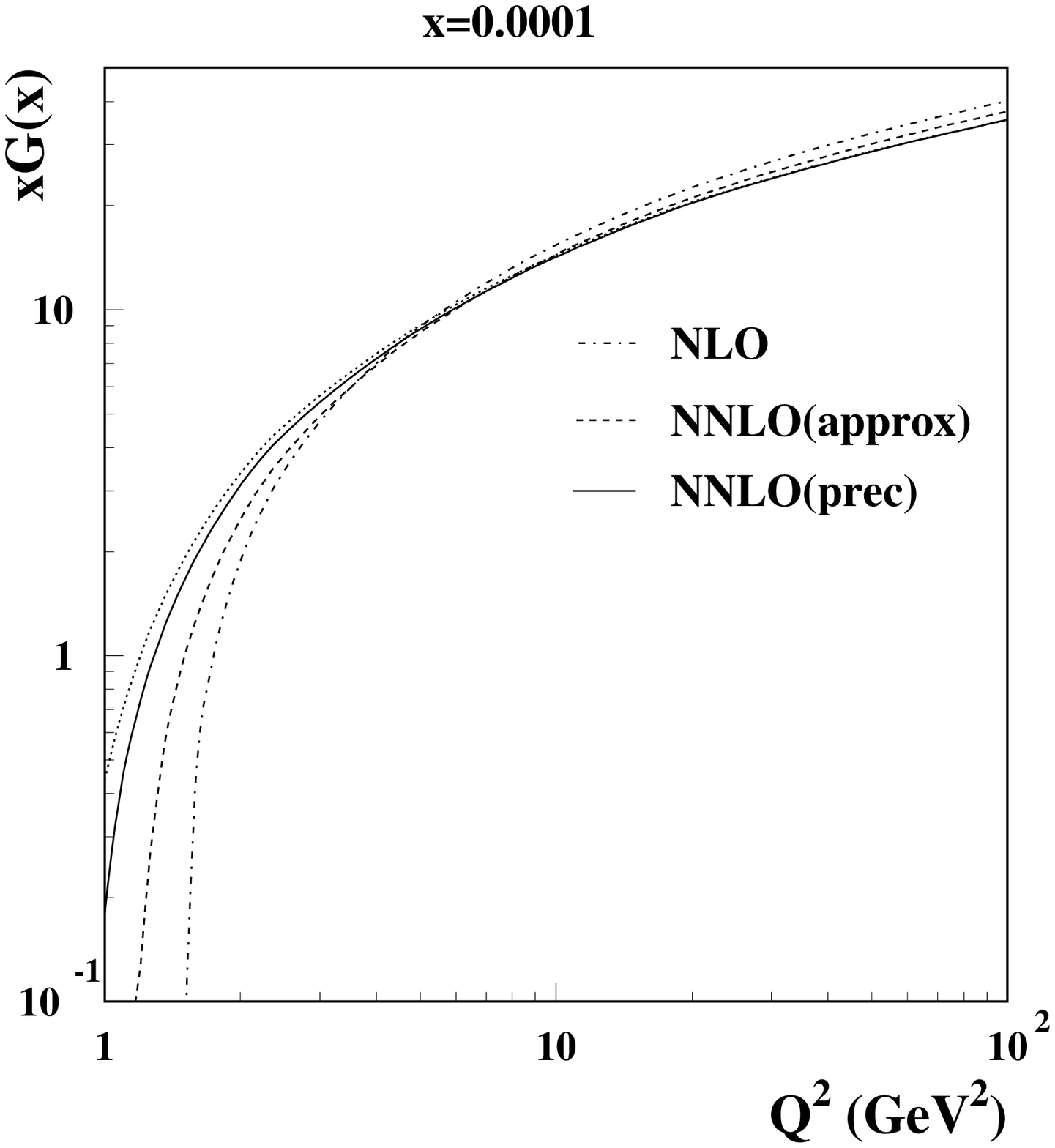,width=18cm,height=18cm}}
\caption{The gluon distributions obtained in the different variants of the
analysis (solid: the fit with the exact NNLO evolution; dashes: 
the fit with approximate NNLO evolution; dots: the approximate NNLO 
gluons evolved with the exact NNLO kernel; dashed-dots: the NLO fit).} 
\label{fig:gluon}
\end{figure}

Parameters of the PDFs obtained in the NNLO fit with their errors due 
to statistical and 
systematical uncertainties in the data are given in Table~\ref{tab:pdfpars}.
These PDFs are comparable to ones of Ref.~\cite{Alekhin:2002fv} extracted using
the approximate NNLO kernel within the errors due to 
the NNLO kernel uncertainty estimated in Ref.~\cite{vanNeerven:2000wp}.
However, at small $x$ and $Q$, where the NNLO corrections 
are enhanced, impact of the new calculations is non-negligible. 
With the exact NNLO corrections the QCD evolution of the gluon
distribution at small $x$ gets weaker and as a result
at small $x/Q$ the gluon distribution obtained using 
the precise NNLO kernel is quite different 
from the approximate one. In particular, while the 
approximate NNLO gluon distribution is negative at 
$Q^2\lesssim1.3~{\rm GeV}^2$, the 
precise one remains positive even below $Q^2=1~{\rm GeV}^2$ 
(see Fig.\ref{fig:gluon}). For the NLO case the positivity of gluons 
at small $x/Q$ is even worse than for the 
approximate NNLO case due to the approximate 
NNLO corrections dampen the gluon evolution at small $x$ too, therefore 
the account of the NNLO corrections is crucial in this respect.
(cf. discussion of Ref.~\cite{Huston:2005jm}).
Positivity of the PDFs is not mandatory beyond the leading order, 
however it 
allows probabilistic interpretation of the parton model and facilitates   
modeling of the soft processes, such as underlying events in the 
hadron-hadron collisions at high energies. 
The change of gluon distribution at small $x/Q$ 
as compared to the fit with approximate NNLO evolution  
is rather due the change in evolution kernel than due to 
shift in the fitted parameters of PDFs. 
This is clear from comparison of the exact NNLO gluon distribution
to one obtained from the approximate NNLO fit and evolved to low $Q$ 
using the exact NNLO kernel (see Fig.\ref{fig:gluon}).
In the vicinity of crossover of the gluon distribution 
to the negative values   
its relative change due to variation of the evolution kernel 
is quite big and therefore further fixation of the kernel at small $x$
discussed in Ref.~\cite{Altarelli:2003hk} 
can be substantial for the low-$Q$ limit of PDFs. 

For the higher-mass kinematics at LHC 
numerical impact of the NNLO kernel update is not dramatic.
The change in the Higgs and $W/Z$ bosons  
production cross sections due to more precise definition of the NNLO 
PDFs is comparable to the errors coming from the PDFs uncertainties.
The NNLO predictions for the 
longitudinal deep-inelastic-scattering structure function 
$F_{\rm L}$ at $x\sim10^{-5}$ measured by the H1 
collaboration~\cite{Lobodzinska:2003yd}
also does not change too much since 
they are given by the Mellin convolution of PDFs
with the coefficient functions and are defined by 
the gluon distribution at relatively big values of $x$.
The obtained value of the strong coupling constant 
$$
\alpha^{\rm NNLO}_{\rm s}(M_{\rm Z})=0.1141\pm 0.0014~({\rm stat+syst}),
$$ 
is in fair agreement to 
$\alpha^{\rm NNLO}_{\rm s}(M_{\rm Z})=0.1143\pm 0.0014~({\rm stat+syst})$
obtained in the fit of Ref.~\cite{Alekhin:2002fv}
with the approximate 
NNLO kernel and to the results of the exact NNLO analysis of the non-singlet
DIS data~\cite{Blumlein:2004ip}.

\begin{table}
\caption{The PDFs parameters and $\chi^2/{\rm NDP}$ obtained in the fit.
The errors in parameters are obtained by propagation of the 
statistical and systematical errors in data.}
\begin{center}
\begin{tabular}{lcc}   
Valence quarks:&&        \\ \hline
            &$a_{\rm u}$&$0.724\pm0.027$\\
       &$b_{\rm u}$&$4.0194\pm0.050$\\
       &$\gamma_2^{\rm u}$&$1.04\pm0.35$\\
       &$a_{\rm d}$&$0.775\pm0.073$\\
       &$b_{\rm d}$&$5.15\pm0.15$\\
Gluon:&&        \\ \hline
&$a_G$&$-0.118\pm0.021$\\
       &$b_{\rm G}$&$9.6\pm1.2$\\
       &$\gamma_1^{\rm G}$&$-3.83\pm0.51$\\
       &$\gamma_2^{\rm G}$&$8.4\pm1.7$\\
Sea quarks:&&        \\ \hline
&$A_{\rm S}$&$0.1586\pm0.0089$\\ 
&$a_{\rm s}$&$-0.2094\pm0.0044$\\
       &$b_{\rm sd}$&$5.6\pm1.2$\\ 
       &$\eta_{\rm u}$&$1.12\pm0.11$\\ 
       &$b_{\rm su}$&$10.39\pm0.88$\\  \hline
$\chi^2/{\rm NDP}$&& 2534/2274 \\ \hline
\end{tabular}
\end{center}
\normalsize
\label{tab:pdfpars}
\end{table}

The errors in PDFs parameters given in Table~\ref{tab:pdfpars} 
are calculated as a propagation of the experimental errors
for the data points used in the fit. We calculate these errors using 
the covariance matrix method~\cite{Alekhin:2000es} taking into account 
statistical and systematic errors in data and correlations of the latter
as well. We also take into account the theoretical errors due to possible 
variations of the 
strange suppression factor $\eta_{\rm s}$ and the $c$-quark mass $m_{\rm c}$.
For this purpose we re-calculate the error matrix 
for the PDFs parameters with $\eta_{\rm s}$ and $m_{\rm c}$ released.
Since the parameters $\eta_{\rm s}$ and $m_{\rm c}$ are not constraint by the 
charged-leptons inclusive DIS data we confine their variation 
adding to the data sample two ``measurements'': $\eta_{\rm s}=0.42\pm0.1$ and  
$m_{\rm c}=1.5\pm0.25~{\rm GeV}$ with the errors in these measurements 
representing our current understanding of the 
uncertainties in these parameters.
In this approach the theoretical errors are included into the 
total error in PDFs and their correlations with other sources of the PDFs 
uncertainties are automatically taken into account.
The NNLO PDFs grid for the range of $Q^2=0.8\div2\cdot 10^8~{\rm GeV}^2$ and 
$x=10^{-7}\div 1$ with the total uncertainties in PDFs supplied is available
directly\footnote{http://sirius.ihep.su/$\tilde{~}$alekhin/pdfa02/}
and through the LHAPDF library\footnote{http://durpdg.dur.ac.uk/lhapdf/}. 
The LO and NLO PDFs grids are also supplied to provide a tool for 
checking sensitivity of different calculation to the QCD order of PDFs.

\begin{figure}
\centerline{\epsfig{file=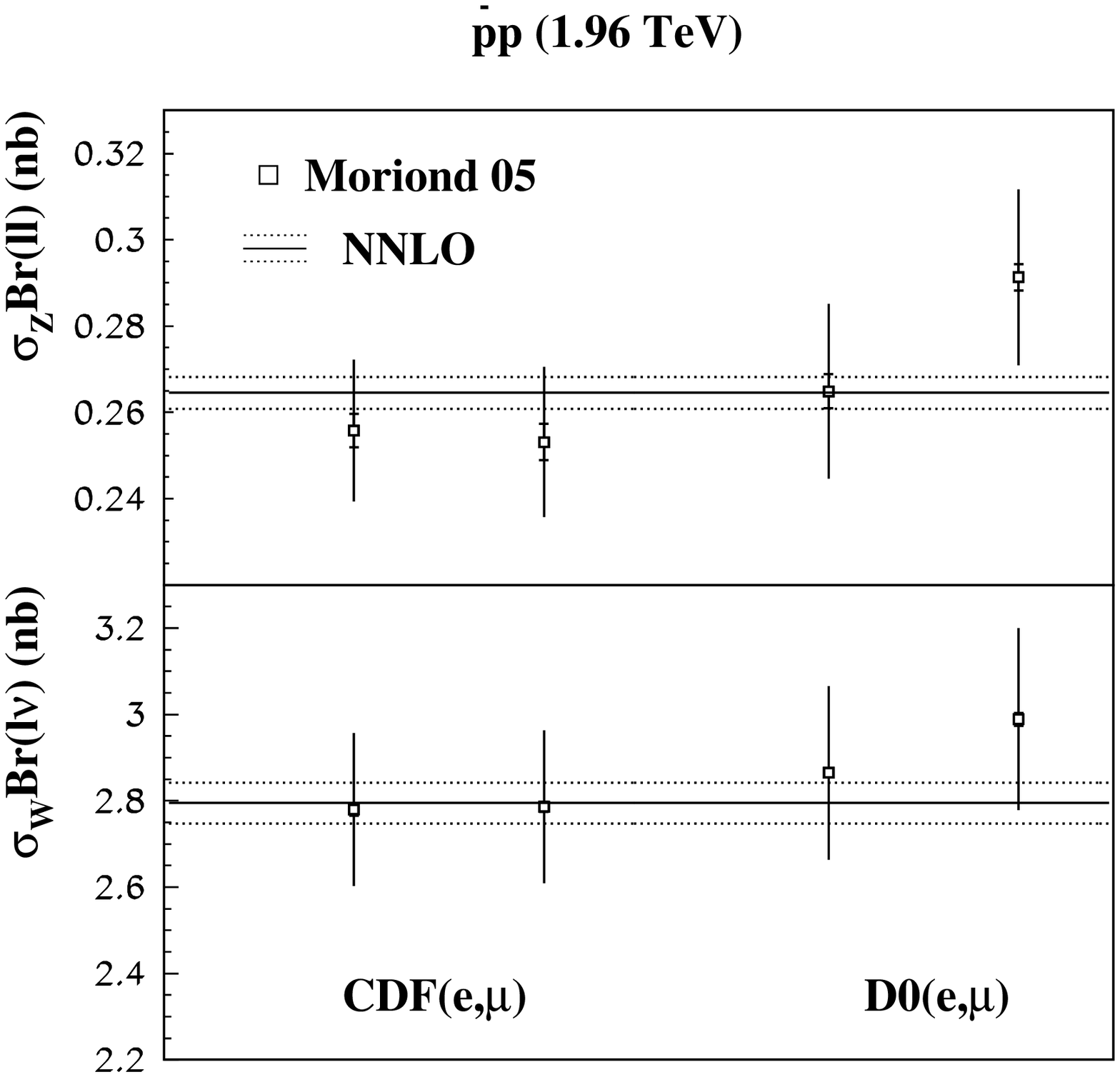,width=18cm,height=18cm}}
\caption{The NNLO calculation of the $W/Z$ rates for Run II at Fermilab 
compared to the data. The dotted lines give the uncertainty in 
calulations due to errors in PDFs; the error bars of the data points give 
the total error including one due to the luminosity uncertainty. 
The branching ratios of the $W/Z$ leptonic decays  
$BR(W \rightarrow l\nu)=0.107$ and $BR(Z \rightarrow ll)=0.034$
were applied.} 
\label{fig:wz}
\end{figure}

The NNLO inclusive rates for the 
intermediate boson production at the FNAL 
$\overline{p}p$ collider and the LHC
calculated using this grid and the code of Ref.~\cite{Hamberg:1990np}
with corrections of Ref.~\cite{Harlander:2002wh}
are given in Table~\ref{tab:wz}. 
The masses and widths of the $W/Z$ bosons   
were set as $M_W=80.425~{\rm GeV}$, $M_Z=91.188~{\rm GeV}$,
$\Gamma_W=2.124~{\rm GeV}$, $\Gamma_Z=2.495~{\rm GeV}$,
squared sine of the Weinberg angle $x_W(M_{\rm Z})=0.2312$, 
squared cosine of the Cabibbo angle $c_C=0.9498$~\cite{Eidelman:2004wy}.
The errors quoted in Table~\ref{tab:wz} 
are due to the total uncertainty in PDFs including the theoretical errors 
considered. The calculations are in agreement to the latest Run II results 
of Ref.~\cite{Bellavance:2005rg}
within the errors (see Fig.~\ref{fig:wz}). The errors in the data of Run II 
are bigger than one in the calculations therefore the latter 
can be used for better 
calibration of the luminosity, which gives main contribution to the 
measurements error. 

\begin{table}
\caption{The NNLO inclusive rates (in nb) 
for the intermediate bosons production in the hadron-hadron collisions.
The errors are due to the total PDFs uncertainties.}
\begin{center}
\begin{tabular}{lcc}   
& $W^{\pm}$ & $Z$ \\ \hline
$\overline{p}p~(1.96~{\rm TeV})$ & $26.11\pm0.44$ & $7.78\pm0.11$ \\
$pp~(14~{\rm TeV})$ & $197.0\pm5.3$ & $57.7\pm1.5$ \\ \hline
\end{tabular}
\end{center}
\normalsize
\label{tab:wz}
\end{table}

In summary, we provide update of the analysis of the world DIS 
inclusive data on the proton/deuteron targets with full account of the 
NNLO QCD corrections including the recent calculations of the 
exact NNLO evolution
kernel. The value of $\alpha_{\rm s}$ is in fair agreement to the earlier 
version of the fit based on the approximate NNLO kernels.  
With the exact NNLO corrections applied
we observe improvement in the positivity 
of the gluon distributions extrapolated to small $x$ and $Q$: Now we have 
gluons positive up to $Q=1~{\rm GeV}$, i.e. throughout kinematical region 
where the parton model is applicable.
The NNLO $W/Z$-bosons rates calculated using the PDFs obtained  
are in agreement with the recent Run II results
and can be used for better calibration of the Fermilab experiments
in view of the uncertainty 
in the calculations due to PDFs are smaller than the experimental ones.
Since these PDFs are extracted from the data for one single process 
they can be used for the quantitative studies of the PDFs universality
that is advantage as compared to ones determined from the global fits. 

{\bf Acknowledgments}

I am indebted to S.Forte, S.Moch, R.Petti, and A.Vogt 
for stimulating discussions and S. Kulagin for development of the Mathematica 
package for access to the PDFs grid.
The work was supported by the RFBR grant 03-02-17177 and 
by the Russian Ministry of Science grant NSh 1695.2003.2.

\end{document}